# Nano-engineering the evolution of skyrmion crystal in synthetic antiferromagnets

Mangyuan Ma[1,†], Ke Huang[2,†], Yong Li[1], Sihua Li[2], Qiyuan Feng[3], Calvin Ching Ian Ang[2], Tianli Jin[2], Yalin Lu[3], Qingyou Lu[3], Wen Siang Lew[2,*], Fusheng Ma[1,*], X. Renshaw Wang[2,4,*]

[1]*Jiangsu Key Laboratory of Opto-Electronic Technology, Center for Quantum Transport and Thermal Energy Science, School of Physics and Technology, Nanjing Normal University, Nanjing, 210046, China*
[2]*Division of Physics and Applied Physics, School of Physical and Mathematical Sciences, Nanyang Technological University, 21 Nanyang Link, Singapore 637371, Singapore*
[3]*Anhui Province Key Laboratory of Condensed, Matter Physics at Extreme Conditions, High Magnetic Field Laboratory, Chinese Academy of Sciences, Hefei 230031, China*
[4]*School of Electrical and Electronic Engineering, Nanyang Technological University, 50 Nanyang Ave, 639798, Singapore*

[†]These authors contributed equally
* phymafs@njnu.edu.cn, wensiang@ntu.edu.sg, & renshaw@ntu.edu.sg

## Abstract

The evolution of skyrmion crystal encapsulates skyrmion's critical behaviors, such as nucleation, deformation and annihilation. Here, we achieve a tunable evolution of artificial skyrmion crystal in nanostructured synthetic antiferromagnet multilayers, which are comprised of perpendicular magnetic multilayers and nanopatterned arrays of magnetic nanodots. The out-of-plane magnetization hysteresis loops and first-order reversal curves show that the nucleation and annihilation of the artificial skyrmion can be controlled by tuning the diameter of and spacing between the nanodots. Moreover, when the bottom layer thickness increases, the annihilation of skyrmion shifts from evolving into a ferromagnetic spin texture to evolving into an antiferromagnetic spin texture. Most significantly, non-volatile multiple states are realized at zero magnetic field *via* controlling the proportion of the annihilated skyrmions in the skyrmion crystal. Our results demonstrate the tunability and flexibility of the artificial skyrmion platform, providing a promising route to achieve skyrmion-based multistate devices, such as neuromorphic spintronic devices.

Ever since the room temperature skyrmions were realized in ferromagnet/heavy metal heterostructures[1–6] with the Dzyaloshinskii-Moriya interaction (DMI),[7,8] the topological behaviors of skyrmions exhibited great potentials in spintronics.[9–12] A recent trend in the investigation of the research field is incorporating skyrmions into advanced artificial structures for both fundamental studies and spintronic applications, such as racetrack memory,[13–15] logic devices[16,17] and skyrmion magnonic crystals.[18–22]

One promising artificial structure is the nano-patterned multilayers.[23–28] Structurally, these nanostructures contain periodic nanodots with tunable sizes, thicknesses and shapes, through which additional tunability can be introduced to the skyrmion system. In these structures, skyrmions nucleate right under these periodic nanodots and inherit the spatial periodicity of the nanodots.[24] The crystal-like group formed by these periodic skyrmions is referred to as skyrmion lattice[26,27] or skyrmion crystal.[24,25,29] Recently, the idea of endowing the skyrmion array with an artificial crystal order attracts much research attention and expands the spintronic research frontier.[18,19] However, to enhance and exploit the functionalities of the skyrmion crystals, introducing engineered spin textures into the skyrmion crystals is desirable yet elusive.

To construct a skyrmion crystal, there are typically two types of skyrmions, namely conventional skyrmion induced by DMI and artificial skyrmion induced by interlayer interaction.[24–27] In particular, the artificial skyrmion crystal gains more functionalities due to the extra tunability from the interface.[30–36] Previous works have demonstrated that artificial skyrmion crystal can be formed *via* the interfacial ferromagnetic coupling between the top magnetic nanodots and the bottom multilayer with perpendicular magnetic anisotropy,[30–33] or *via* the exchange bias in the ferromagnet/antiferromagnet multilayers.[34–36] Recently antiferromagnetic interlayer exchange coupling (AFM-IEC) in synthetic antiferromagnet (SAF) has drawn much attention due to its capabilities in stabilizing skyrmions or generating functional spin textures.[37–42] Therefore, engineering the artificial skyrmion crystal in SAF nanostructures might introduce controlled spin textures, permit unseen functionalities and boost the control of skyrmions.

In this work, we engineer the SAF into nanostructures, which contain the top nanodots and the continuous bottom layer with varied geometries. A periodically-ordered skyrmion crystal is generated in these nanostructures. The out-of-plane magnetization hysteresis ($m_z$-$H$) loops and first-order reversal curves (FORCs) reveal that the evolutions of the skyrmion crystal, which are nucleation and annihilation, strongly depends on the thicknesses of the bottom magnetic layer and the diameters of the top nanodots. Especially, by increasing the thickness of the bottom layer, skyrmions can be switched from evolving into a ferromagnetic to an antiferromagnetic spin texture. We also find that even with part of the skyrmion crystal evolving into other spin textures, the rest is still stable. Based on this feature, we achieve non-volatile multiple states, characterized by multiple levels of Hall resistance.

**Results and discussion**

Figure 1a shows the schematic structure of the nanostructured SAF multilayers Ta(4)/Pt(4)/[Pt(0.6)/Co(0.6)]$_N$/Ru(0.9)/[Co(0.6)/Pt(0.6)]$_4$/Ta(4), where the numbers in parentheses are the nominal layer thickness in nanometer. Detailed information of sample preparation can be found in the Methods section. The spacing layer Ru provides an AFM-IEC between the bottom [Pt/Co]$_N$ and top [Co/Pt]$_4$ layers. The top Ta layer is used to prevent oxidation, and the bottom Ta and Pt layers work as seeding layers. The bottom [Pt/Co]$_N$ layer has a repeating number $N$ ($N$ = 2 or 4). Note that the repeating number $N$ = 2 or 4 ensures a single-magnetic-domain state in the bottom [Pt/Co]$_N$ layer. When $N$ takes a larger value, the domain in the magnetic layer may exhibit a multi-domain state,[43–45] which is unfavorable in this work. The top [Co/Pt]$_4$ layer was etched into circular nanodots with diameters ($d$) varied from 100 to 600 nm. In all samples, the spacing between the nanodots ($S$) is equal to $d$. We prepared the nanostructured SAF multilayers with two types of nanodots distributions, namely (i) two-dimensional arrangement of nanodots for magneto-optical Kerr effect (MOKE) measurement (see scanning electron microscopy (SEM) image of $d$ = 600 nm at the bottom left of Figure 1a) and (ii) a finite number $Q$ of nanodots for Hall resistance measurement (see SEM image of $d$ = 400 nm at the bottom right of Figure 1a for $Q$ = 9). Note that $Q$ = 0 is the sample with continuous ferromagnetic multilayers [Pt/Co]$_N$, and it serves as a reference sample. The schematics of the samples for MOKE and Hall resistance measurements can be found in Section II of Supporting Information (SI). We define the upward and downward directions as the positive and negative directions of the external magnetic field, respectively (Figure 1a).

The $m_z$-$H$ loops of nanostructured SAF multilayers were measured by MOKE. Technical details of the MOKE measurement are presented in the Methods section. Figure 1b,c shows the descending branches of the MOKE-measured $m_z$-$H$ loops of the nanostructured SAF multilayers with $N$ = 2, $d$ = 600 nm and $N$ = 4, $d$ = 600 nm. Both of the $m_z$-$H$ loops exhibit multi-step switching, which corresponds to different magnetization distributions.[46] We define the coercive field ($H_c$) as the magnetic field value at which the $m_z$ crosses zero (see Section III of SI). Additionally, to identify all the magnetization states, we performed magnetic force microscopy (MFM) measurements on the nanostructured SAF multilayers with $N$ = 2, $d$ = 600 nm at various fields (see four typical fields in Figure 1d). Technical details of the MFM measurement are presented in the Methods section. We found that, along the descending branch of the $m_z$-$H$ loop, the nanostructure with $N$ = 2/4, and $d$ = 600 nm sequentially experiences four/five distinct magnetization states. In order to elaborate the different magnetic transitions clearly for $N$ = 2/4, we presented six examples of cross-section views of magnetization distribution at representative magnetic fields, as shown in insets of Figure 1b,c. The corresponding regions and MFM images are labelled in Figure 1b,c, and Figure 1d, respectively.

We now first focus on the process of magnetization switching of the $N$ = 2, $d$ = 600 nm samples. When the external field is larger than the saturation field ($H_{sat}$), the magnetization of the top and bottom layers are aligned along the direction of the external field (inset 1 of Figure 1b,d for $H$ = 12 kOe). As the external field decreases towards 0 kOe, the magnetization of the top nanodots is reversed by the AFM-IEC (inset 2 of Figure 1b,d for $H$ = 0.8 kOe). Because the magnetizations of the top nanodots and the bottom continuous layer are antiparallel, we denote this magnetization configuration as an antiferromagnetic spin texture (AST) state. When the negative external field is greater than the coercive field ($H_c$), the magnetization of the dot-uncovered region of the bottom layer is reversed downward. The magnetization of the dot-covered region remains antiparallel to that of the top nanodots because of the AFM-IEC. Considering solely the bottom continuous layer, a skyrmion-like magnetization configuration nucleates (inset 3 of Figure 1b,d for $H$ = -2.6 kOe), which is referred to as a skyrmion (SK) state. Further increasing the external field negatively will gradually offset the AFM-IEC protection, resulting in a shrinking annihilation of the skyrmion (inset 4 of Figure 1b,d for $H$ = -2.6 kOe). Because the negative external field is larger than $H_{sat}$, the magnetization of the bottom layer is fully reversed downward, and this state is referred to as a ferromagnetic spin texture (FST) state (inset 5 of Figure 1b,d for $H$ = -12 kOe).

Similarly, Figure 1c shows the magnetization evolution of nanostructured SAF multilayers with $N$ = 4, $d$ = 600 nm. In particular, an additional AST state with a rise of $m_z$ occurs after the annihilation of the artificial skyrmions (inset 6 of Figure 1c). In comparison to the $N$ = 2 nanostructure, the additional AST state is induced by a larger AFM-IEC in the $N$ = 4 nanostructure. The difference of AFM-IEC between the $N$ = 2 and $N$ = 4 nanostructures is confirmed by comparing the $m_z$-$H$ loops of continuous SAF multilayers with different $N$ (Section IV of SI).

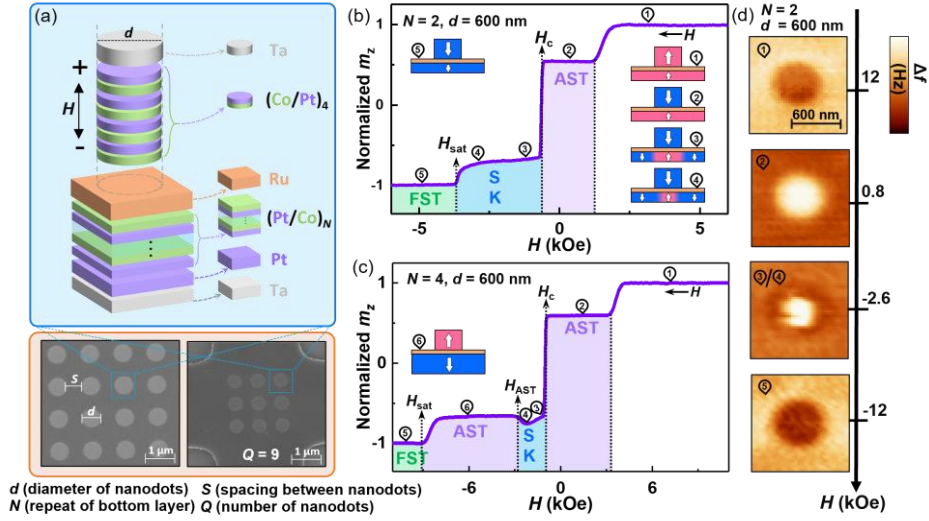

Figure 1. Magnetization switching processes of nanostructured SAF multilayers. (a) Schematic illustration of nanostructured SAF multilayers. Bottom left and bottom right images are the SEM images of a 2D arrangement of the top nanodots with diameter $d$ = 600 nm and a Hall bar with 9 top nanodots with diameter $d$ = 400 nm, respectively. (b,c) MOKE-measured descending branches of $m_z$-$H$ loops of nanostructured SAF multilayers with $d$ = 600 nm for $N$ = 2 (b) and 4 (c). Insets are cross-section views of magnetization distributions at selected fields. (d) MFM images of nanostructured SAF multilayers with $N$ = 2 and $d$ = 600 nm at typical fields. The fields are selected at positive FST, AST, SK, and negative FST states. The MFM contrast, which is caused by the resonant frequency shift ($\Delta f$) of the MFM cantilever, shows a dark/bright contrast when the interaction between the MFM tip and the local magnetization of the sample is attractive/repulsive.

In order to investigate the effect of $d$ on the nucleation and shrinking annihilation of skyrmions in nanostructured SAF multilayers, $d$ was varied from 100 to 600 nm. Figure 2a,b shows the fractions of the MOKE-measured $m_z$-$H$ loops with various $N$ and $d$ as well as of the reference samples with continuous ferromagnetic [Pt/Co]$_2$ and [Pt/Co]$_4$ multilayers. The full $m_z$-$H$ loops are shown in Section V of SI. The critical external field of skyrmion nucleation ($H_c$) decreases and gradually approaches the $H_c$ of the reference samples, when $d$ increases from 100 to 600 nm for both $N$ = 2 and 4 samples. On the contrary, the critical external fields of skyrmion shrinking annihilation, which are $H_{sat}$ for $N$ = 2 and $H_{AST}$ for $N$ = 4, increase as $d$ increase. The different dependence of $H_c$, $H_{sat}$ and $H_{AST}$ against $d$ can be understood from the different switching scenarios of the bottom continuous [Pt/Co]$_N$ layer. Taking $N$ = 2 as an example, Figure 2c schematically shows that the nanostructured SAF multilayers can be divided into three regions, namely (A) the dot-covered region, (B) the dot-non-covered region, and (C) the top nanodots. When the external field negatively increases, the magnetization transitions of the AST-SK and the SK-FST manifest as the switching of regions B and A, respectively (Figure 2d). The AST-SK magnetization transition is realized by the reversal of region B, which is collectively determined by three interactions: (i) the coercivity field ($H_{c\text{-B}}$) of the continuous bottom [Pt/Co]$_2$ layer, (ii) the ferromagnetic exchange interaction field ($H_{\text{ex-AB}}$) at the interface between regions A and B, and (iii) the external field ($H$). Figure 2d shows that, during the skyrmion nucleation, the magnetization of region B is reversed when $H$ is larger than the AST-SK transition field ($H_c$). The value of $H_c$ is decided by the summation of $H_{\text{ex-AB}}$ and $H_{c\text{-B}}$:

$$|H_c| = |H_{\text{ex-AB}}| + |H_{c\text{-B}}|, \tag{1}$$

where both $H_{\text{ex-AB}}$ and $H_{c\text{-B}}$ protect the region B from being reversed by $H$. However, when $H = H_c$, the magnetization of region B reverse downward, the magnetization of region A is retained by the AFM-IEC between regions A and C. As a result, region A tends to protect the region B from being reversed by $H_{\text{ex-AB}}$. For a smaller $d$ ($d = S$), the contribution of $H_{\text{ex-AB}}$ to $H_c$ is relatively larger. While $H_{c\text{-B}}$ is dependent on the coercivity field of the continuous [Pt/Co]$_2$ layer but independent of $d$, the inverse dependence of $H_c$ on $d$, therefore, can be understood according to eq 1.

On the other hand, the SK-FST magnetization transition is realized by the reversal of region A. This transition is collectively determined by four parameters: (i) the coercivity field ($H_{c\text{-A}}$) of the continuous bottom [Pt/Co]$_2$ layer, (ii) the ferromagnetic exchange interaction field ($H_{\text{ex-BA}}$) at the interface between regions A and B, (iii) the AFM-IEC field ($H_{\text{ex-CA}}$) between regions A and C, and (iv) $H$. Figure 2d shows that the skyrmion shrinks and annihilates where the magnetization of region A could be reversed and when $H$ is larger than the SK-FST transition field $H_{\text{sat}}$ of the SK-FST transition. The value of $H_{\text{sat}}$ is determined by

$$|H_{\text{sat}}| = |H_{\text{ex-CA}}| + |H_{c\text{-A}}| - |H_{\text{ex-BA}}|, \tag{2}$$

where both $H_{\text{ex-CA}}$ and $H_{c\text{-A}}$ protect region A from being reversed by $H$. $H_{\text{ex-CA}}$ and $H_{c\text{-A}}$ are determined by the AFM-IEC strength and the coercivity field of the continuous [Pt/Co]$_2$ layer, respectively. But, both are independent of $d$. In contrast, region B tends to reverse region A *via* $H_{\text{ex-BA}}$. And, for a smaller $d$, the $H_{ex-BA}$ is relatively larger. Hence, the dependence of $H_{sat}$ on $d$ can be understood according to eq 2, and a smaller $d$ results in a smaller $H_{sat}$. Similarly, the dependence of $H_c$ and $H_{AST}$ on $d$ for $N = 4$ can also be understood by considering the field-dependent reversal of regions A, B and C. Therefore, the field ranges of the SK state, which are $|H_{sat} - H_c|$ for $N = 2$ and $|H_{AST} - H_c|$ for $N = 4$, can be manipulated by tailoring $d$, and the tunability further enables the flexibility of the nanostructured SAF multilayers in skyrmionic and magnonic applications.

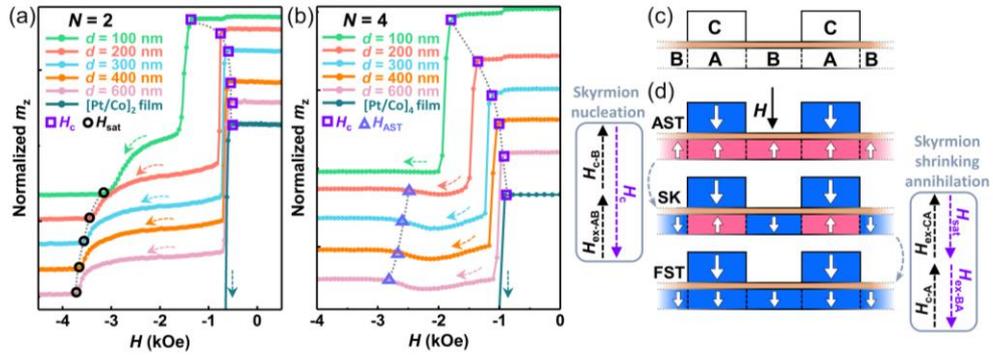

Figure 2. Nucleation and shrinking annihilation of artificial skyrmions in nanostructured SAF multilayers. (a,b) Fractions of the descending branches of MOKE-measured $m_z$-$H$ loops of nanostructured SAF multilayers with top nanodots of various diameters for $N = 2$ (a) and 4 (b). Open squares, circles and triangles represent $H_c$, $H_{sat}$ and $H_{AST}$, respectively. (c) Cross-section view of nanostructured SAF multilayers with $N = 2$. Regions A, B and C are dot-covered region, dot-non-covered region and top nanodot region, respectively. (d) Magnetization switching of regions B and A during the AST-SK and the SK-FST transition processes.

Furthermore, in contrast to the skyrmion's shrinking annihilation in form of SK-FST and SK-AST transitions, skyrmions can also annihilate in a form of expansion. To investigate the skyrmion-expansion-induced magnetization transition, we measured minor $m_z$-$H$ loops using MOKE, *i.e.*, $H_{sat}$-$H_R$-$H_{sat}$ loops, by choosing a reversal field $H_R < H_{sat}$. Figure 3a, b shows minor loops of $d = 300$ nm samples with $N = 2$ and 4, respectively. Each figure contains two minor loops with $H_R$ located in SK state ($H_R = H_{R1}$) and FST/AST state ($H_R = H_{R2}$). We observed that the magnetization $m_z(H, H_R)$ is highly dependent on the selection of $H_R$, when the $H$ increases from $H_R$ toward $H_{sat}$. For instance, the corresponding cross-sections of three loops with different $H_R$ at $H = 2$ kOe are indicated by vertical dashed lines in Figure 3a,b. This difference of magnetization distribution originates from the $H_R$-dependent magnetization transition. When $H_R = H_{R1}$, the SK-AST transition occurs in both $N = 2$ and 4 samples together with a type of skyrmions annihilation in a form of expansion. When $H_R = H_{R2}$, the minor loops coincide with the ascending branches of major loops. Skyrmion nucleation occurs when the minor loops cross positive $H_c$.

Because the skyrmion-expansion-induced magnetization transition depends on the value of $H_R$, a systematic investigation of $m_z(H, H_R)$ is necessary. Here we used the FORC technique for the detailed exploration. FORC technique was used in the MOKE measurement by changing $H_R$ with an interval of $\Delta H_R$. The FORCs, which is comprised of a large number of "$H_{sat}$-$H_R$-$H_{sat}$" minor loops,[47–49] have been widely used to study the distribution of switching fields, the interaction fields between neighboring domains, and the irreversibility of the magnetization switching mechanism.[50–52] Details of FORCs and calculation of FORC diagram are described in Section VI of SI. Figure 3c,d shows the fractions of MOKE-measured FORCs in the form of contour plots of $m_z(H, H_R)$ with $d$ varied from 100 to 600 nm for $N = 2$ and 4 samples. The corresponding full FORCs and corresponding FORC diagrams can be found in Figure S6 of SI. Depending on the values of $H_R$, there are two left-right boundaries in the contour plots of $m_z(H, H_R)$, and they are indicated as boundary 1 and 2 in Figure 3c,d. Now, we use the minor loops and the contour plots of $m_z(H, H_R)$ for $N = 2$, $d = 300$ nm (Figure 3a,c) to elaborate the origins behind the presence of boundaries. When $H_R = H_{R1}$, boundary 1 corresponds to the SK-AST transition (insets of Figure 3c). As for $H_R = H_{R2}$, boundary 2 corresponds to the AST-SK transition (insets of Figure 3c). In addition, the position of boundary 1/2 decreases/increases to low/high $H$ when $d$ decreases from 600 to 100 nm. This dependence of boundary position indicates that the skyrmions in larger $d$ samples can expand continuously to the positive field region. On one hand, skyrmion can stabilize under zero field, when $H_R$ locates at the SK state. On the other hand, the skyrmions can nucleate more easily when $H_R$ locates at the FST state for the samples with larger $d$.

Figure 3d shows the contour plots of $m_z(H, H_R)$ for $N = 4$ sample with two left-right boundaries. Similar to the boundaries of $N = 2$ sample, the boundary 1/2 corresponding to the SK-AST and the AST-SK transition for $N = 4$ sample by considering the minor loops in Figure 3b. The boundary 1/2 of $N = 4$ sample moves similarly as that of $N = 2$ sample. Note that no boundary 1 occurs in $N = 4$ and $d = 100$ nm sample, indicating that a SK state hardly exists in this case. This is consistent with our experimental results of $m_z$-$H$ loop of $d = 100$ nm in Figure 2b.

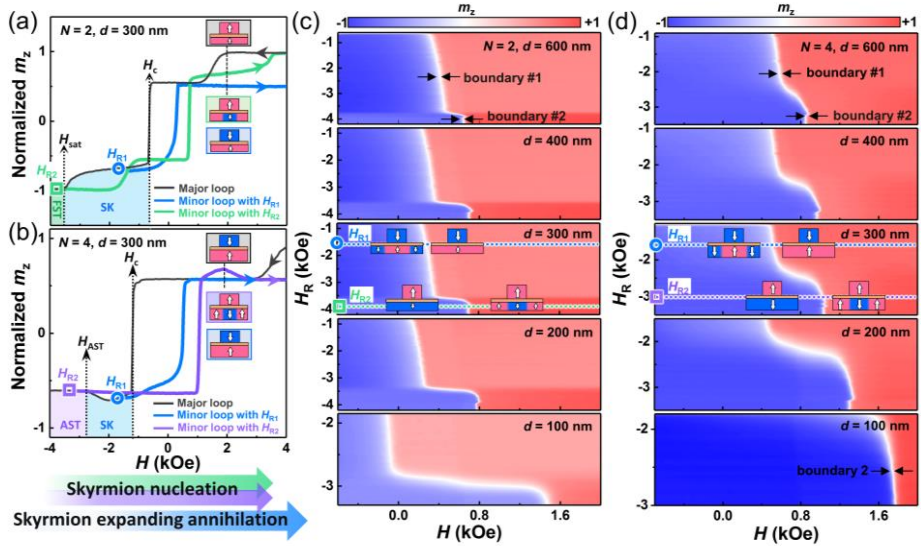

Figure 3. MOKE-measured FORCs of the skyrmion nucleation and skyrmion expanding annihilation. (a,b) Fractions of major and minor $m_z$-$H$ loops of nanostructured SAF multilayers with $d = 300$ nm for (a) $N = 2$ and (b) $N = 4$. Black, blue and green (purple) solid lines represent the major loop, minor loop with $H_{R1}$ selected at the SK state, and minor loop with $H_{R2}$ selected at the FST (AST) state, respectively. Insets are the schematic cross-sections of the three loops at $H = 2$ kOe. (c,d) Fractions of FORCs presented in the form of contour plots of $m_z(H, H_R)$ with various $d$ in (c) $N = 2$ and (d) $N = 4$. Horizontal dash lines represent the evolutions of $m_z(H, H_R)$ along with the minor loops with $H_{R1}$ (open circle) and $H_{R2}$ (open square). Insets represent the SK-AST and the AST-SK transitions along with the minor loops with $H_R = H_{R1}$ and $H_R = H_{R2}$, respectively.

So far, we have mainly focused on the intrinsic properties of skyrmions in the nanostructured SAF multilayers with various $N$ and $d$. Now, we switch to Hall resistance measurements to investigate the potential applications based on the evolution of skyrmions. Figure 4a shows the evolution of the Hall resistance ($R_{xy}$) of the nanostructured SAF multilayers with fixed $N = 4$ and $d = 400$ nm, but varied $Q = 1$ and 9. In order to reveal the influence of $Q$ on the behaviors of skyrmion, a zoomed-in view of the SK region is presented in Figure 4b. The SK region can be divided into stabilization and annihilation regions, which correspond to the shrinking of the skyrmion and the SK-AST transition, respectively. We found that the annihilation region of skyrmions in $Q = 9$ nanostructure is much larger than that in $Q = 1$ nanostructure, which indicates the nine skyrmions in the $Q = 9$ nanostructure annihilate asynchronously. Based on this phenomenon, we could purposely nucleate or annihilate different numbers of skyrmions to achieve multiple states at zero field. Note that the annihilation region manifests as a rise in the $m_z$-$H$ loop (Figure 1c) but a sharp drop in the Hall resistance hysteresis ($R_{xy}$-$H$) loop. This should be because Hall resistance measurement is more sensitive to the continuous bottom layer and less sensitive to the etched top layer.

To experimentally realize the multiple states base on the different number of skyrmions, we measured the reversal curves with the reversal fields $H_R$ selected at different fields, where there are different numbers of skyrmions. Figure 4c shows a full major curve and three reversal curves with $H_R$ = -1.6 kOe, -2.2 kOe and -2.4 kOe. Figure 4d shows the zoomed-in view of the three reversal curves. Note that we only present the results of the descending branch. Similar states could also be achieved in the ascending branch of the positive field region. As shown in Figure 4d, we obtained four different states at zero field, namely base state, state 1, state 2 and state 3. The base state was realized by sweeping the external field from $H_R$ = -$H_{sat}$ to zero field, and there is no skyrmion at the $H_R$. State 1, 2 and 3 were achieved by sweeping the external field from different $H_R$ to zero field, and there are different numbers of skyrmions at these selected $H_R$. Moreover, we conducted local hysteresis loops sweeping between $H_R$ and zero field for the four states. The value of $R_{xy}$ at zero field is always repeatable (Figure 4d), which indicates that the multiple states base on the different numbers of skyrmions are non-volatile and robust.

We attribute these multiple states to an asynchronous evolution of the skyrmions with the external field. For the nanostructured SAF multilayers with the periodically-arranged nanodots, the skyrmions forms in the bottom layer are also arranged periodically and behave the same as their neighbors. However, the nine skyrmions in the nanostructured SAF multilayers with $Q = 9$ can be divided into three types according to the difference in their positions: (i) four skyrmions at the corner, (ii) four skyrmions at the edge center, and (iii) one skyrmion at the center. The evolutions of the three types of skyrmions with the external field are different, which is confirmed by the micromagnetic simulation in Section VIII of SI (the relevant simulation parameters are given in Section I of SI). Our Hall resistance measurement and simulated results further demonstrate that, if we could introduce additional asymmetry of skyrmions through tailoring the shape, distribution and size of the nanostructure, the nucleation and annihilation of skyrmion can be controlled more accurately.

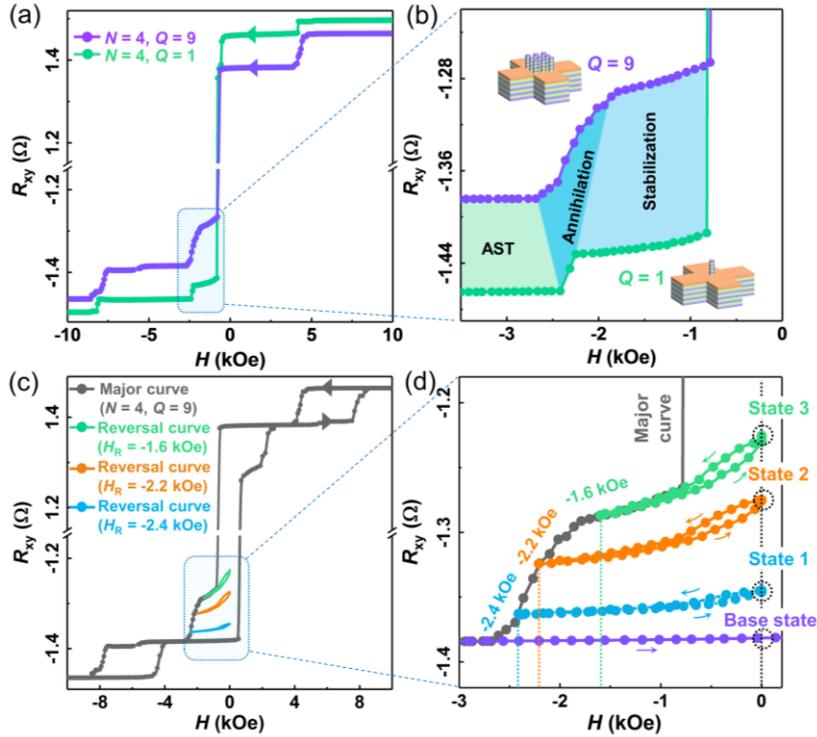

Figure 4. Skyrmion-based non-volatile multiple states. (a) Descending branches of $R_{xy}$-$H$ loops of nanostructured SAF multilayers with $N = 4$, $Q = 1$ and 9. (b) Zoomed-in view of the region indicated by the blue frame in (a). Insets are the schematic illustration of nanostructured SAF multilayers with $Q = 1$ and 9. (c) Major and reversal curves of the nanostructured SAF multilayers with $N = 4$, $Q = 9$. (d) Zoomed-in view of the region indicated by the blue frame in (c). Dash line circles indicate the zero-field multiple states achieved by the reversal curves with different reversal fields: $H_R$ = -1.6, -2.2 and -2.4 kOe.

## Conclusions

In summary, we experimentally investigated the behaviors of skyrmion crystal in the nanostructured SAF multilayers and validated a feasible scheme for the application of the multilayers. The existence of the skyrmion crystal in nanostructured SAF multilayers and its shrinking annihilation, *i.e.*, the SK-FST and SK-AST transitions, were characterized by the MOKE-measured $m_z$-$H$ loops. We also proved that the external field range of the existence of skyrmions could be improved *via* increasing the diameter of top nanodots *d*. Moreover, with the help of the FORC technique, we found that the zero-field stability and field range of existence of skyrmions can be improved by choosing the reversal field at the SK region and tailoring *d*. Last, according to the asynchronous annihilation of skyrmions, we achieved the zero-field non-volatile multiple states. The proposed structure can significantly widen the scope of the skyrmion material, and the multiple states could be applied in the field of data storage and neuromorphic computing.

**Methods**

*Sample Preparation*: The used synthetic antiferromagnetic multilayers in this work, Ta(4)/Pt(4)/[Pt(0.6)/Co(0.6)]$_N$/Ru(0.9)/[Co(0.6)/Pt(0.6)]$_4$/Ta(4), were sequentially deposited over oxidized silicon wafers using a DC magnetron sputtering technique. *N* is 2 or 4 in our work. The numbers in the bracket are the nominal thickness of the corresponding layer. During the deposition, 2.3 × $10^{-3}$ Torr Argon gas was filled in the vacuum chamber with a base pressure of 2 × $10^{-8}$ Torr. The deposition rates were 0.21, 0.14, and 0.10 Å/s for Co, Pt, and Ru, respectively. The multilayers were then spin-coated with Polymethyl methacrylate (PMMA). Electron beam lithography (EBL) was then used to pattern the PMMA layer with circular nanodots. The diameter and spacing of the circular nanodots varied from 100 to 600 nm. Then, Ar$^+$ ion milling was used to remove the part of the sample above Ru but uncovered by circular PMMA nanodots. Before etching the sample, we calibrated the etching rate of each component in the sample based on both deposition time and machine monitoring. During the etching process, we monitored the sample elements and stop the etching immediately after the Ru layer appears. Under the dual operation of time controlling and machine monitoring, we can realize that the etching mainly on the top [Co/Pt]$_4$ layer. Finally, we dissolved the remaining PMMA on the sample and obtained patterned nanodots above Ru. The steps above are used to prepare the samples for MOKE measurements. To prepare the samples for Hall resistance measurements, we further etched the Ru layer and the bottom [Pt/Co]$_N$ layer below Ru into a cross bar shape (length 40 µm and width 3 µm) using EBL and Ar$^+$ ion milling methods. The samples used for MOKE measurements have periodic nanodots, while the samples used for Hall resistance measurements have a single nanodot or 9 nanodots as shown in Figure 1a.

*Magneto-Optical Kerr effect and Hall Resistance Measurements*: The $m_z$-$H$ loops and the FORCs were measured by the MagVision Kerr imaging system. Differential imaging technology and piezoelectric actuators were applied to enhance the detected magnetic signal and to eliminate the influence of sample drift, respectively. Please refer to https://www.vertisis.com.sg/. The Hall resistance measurements were conducted with an AC/DC current source of Keithley 6221 and a nano voltmeter of Keithley 2182A. The current used in all the Hall resistance measurements was a square wave with a peak-to-peak amplitude of 0.2 mA.

*Magnetic Force Microscopy*: MFM measurements were conducted in a home-built system. This system has a superconducting magnet with a maximum magnetic field of 20 T.[53,54] The piezoresistive cantilever of the system is the commercial PRC400 from Hitachi High-Tech Science Corporation. This cantilever has a 42 kHz resonant frequency. The MFM tip is coated with Cr, Fe, and Au films. The thicknesses of Cr, Fe, and Au coating are 5, 50, and 5 nm, respectively. The coercivity and saturation fields of the MFM tip are $H_c \approx 0.25$ kOe and $H_{sat} \approx 2$ kOe, respectively. The MFM uses a built-in phase-locked loop to adjust the scanning process and process signal. During the MFM measurement, the contact mode was first used to obtain a topographic image. According to the topographic image, the sample surface tilting along the fast and slow scan axes can be compensated. Second, the MFM images were measured in a frequency-modulation mode with a tip height of ~100 nm.

**Acknowledgment**

F.M. acknowledges support from the National Natural Science Foundation of China (Grant No. 12074189 and 11704191). X.R.W. acknowledges supports from Academic Research Fund Tier 2 (Grant No. MOE-T2EP50120-0006) from the Singapore Ministry of Education and Agency for Science, Technology, and Research (A*STAR) under its AME IRG grant (Project No. A20E5c0094). Q.L. and Q.F. acknowledge support from the National Key R&D Program of China (Grant Nos. 2017YFA0402903 and 2016YFA0401003), the National Natural Science Foundation of China (Grant No. 51627901, U1932216), Hefei Science Center CAS (Grant 2018HSC-UE014), and the Maintenance and Renovation Project for CAS Major Scientific and Technological Infrastructure (Grant DSS-WXGZ-2019-0011). W.S.L. acknowledges support from an NRF-CRP Grant (No. CRP9-2011-01), a RIE2020 ASTAR AME IAF-ICP Grant (No. I1801E0030), and an ASTAR AME Programmatic Grant (No. A1687b0033). W.S.L. is a member of the SG-SPIN Consortium.

**Supporting Information Available**

Micromagnetic simulation details, schematics of the samples for MOKE and Hall resistance measurements, definition of the coercive field, strength of antiferromagnetic interlayer exchange coupling, full descending branches of $m_z$-$H$ loops of nanostructured SAF multilayers, FORC measurement, analysis of FORC diagram, asynchronous evolution of skyrmions